\documentclass[pdflatex,sn-mathphys-num]{sn-jnl}

\usepackage{graphicx}%
\usepackage{multirow}%
\usepackage{amsmath,amssymb,amsfonts}%
\usepackage{amsthm}%
\usepackage{mathrsfs}%
\usepackage[title]{appendix}%
\usepackage{xcolor}%
\usepackage{textcomp}%
\usepackage{manyfoot}%
\usepackage{booktabs}%
\usepackage{algorithm}%
\usepackage{algorithmicx}%
\usepackage{algpseudocode}%
\usepackage{listings}%
\usepackage{lineno}

\theoremstyle{thmstyleone}%
\theoremstyle{thmstyletwo}%
\theoremstyle{thmstylethree}%
\begin{document}

\title[Rapid Dust Formation in the Early Universe]{Rapid Dust Formation in the Early Universe}

\author*[1]{\fnm{Danial} \sur{Langeroodi}}\email{danial.langeroodi@nbi.ku.dk}

\author[1]{\fnm{Jens} \sur{Hjorth}}

\author[2]{\fnm{Andrea} \sur{Ferrara}}

\author[1]{\fnm{Christa} \sur{Gall}}

\affil[1]{\orgdiv{DARK, Niels Bohr Institute}, \orgname{University of Copenhagen}, \orgaddress{\street{Jagtvej 155A}, \city{2200 Copenhagen}, \country{Denmark}}}

\affil[2]{\orgdiv{Scuola Normale Superiore}, \orgaddress{\street{Piazza dei Cavalieri 7}, \city{50126 Pisa}, \country{Italy}}}

\abstract
{\textbf{
Interstellar dust links the formation of the first stars to the rocky planet we inhabit by playing a pivotal role in the cooling and fragmentation of molecular clouds, and catalyzing the formation of water and organic molecules. Despite its central role, the origin of dust and its formation timescale remain unknown \citep{2011A&ARv..19...43G, Dayal18, Schneider+2024}. Some models favor rapid production in supernova ejecta as the primary origin of dust, while others invoke slower production by evolved asymptotic giant branch stars or grain growth in the interstellar medium (ISM). The dust content of young early-universe galaxies is highly sensitive to the dust formation timescales. Here, we evaluate the dust content of 631 galaxies at $3 < z_{\rm spec} < 14$ based on rest-UV to optical spectroscopy obtained with JWST NIRSpec. We find that dust appears rapidly. Attenuation immediately follows star formation on timescales shorter than $\sim 30$ Myr, favoring dust production by supernovae. The degree of attenuation is $\sim 30$ times lower than expected if the entire supernova dust yield were preserved in the ISM, and had Milky Way-like grain properties. This can be reconciled if the early-universe dust is composed mostly of silicate or grains much larger than those in the Milky Way, and if significant dust destruction or ejection by outflows takes place.
}}

\maketitle

\begin{figure*}
\centering\includegraphics[width = 1.0 \textwidth]{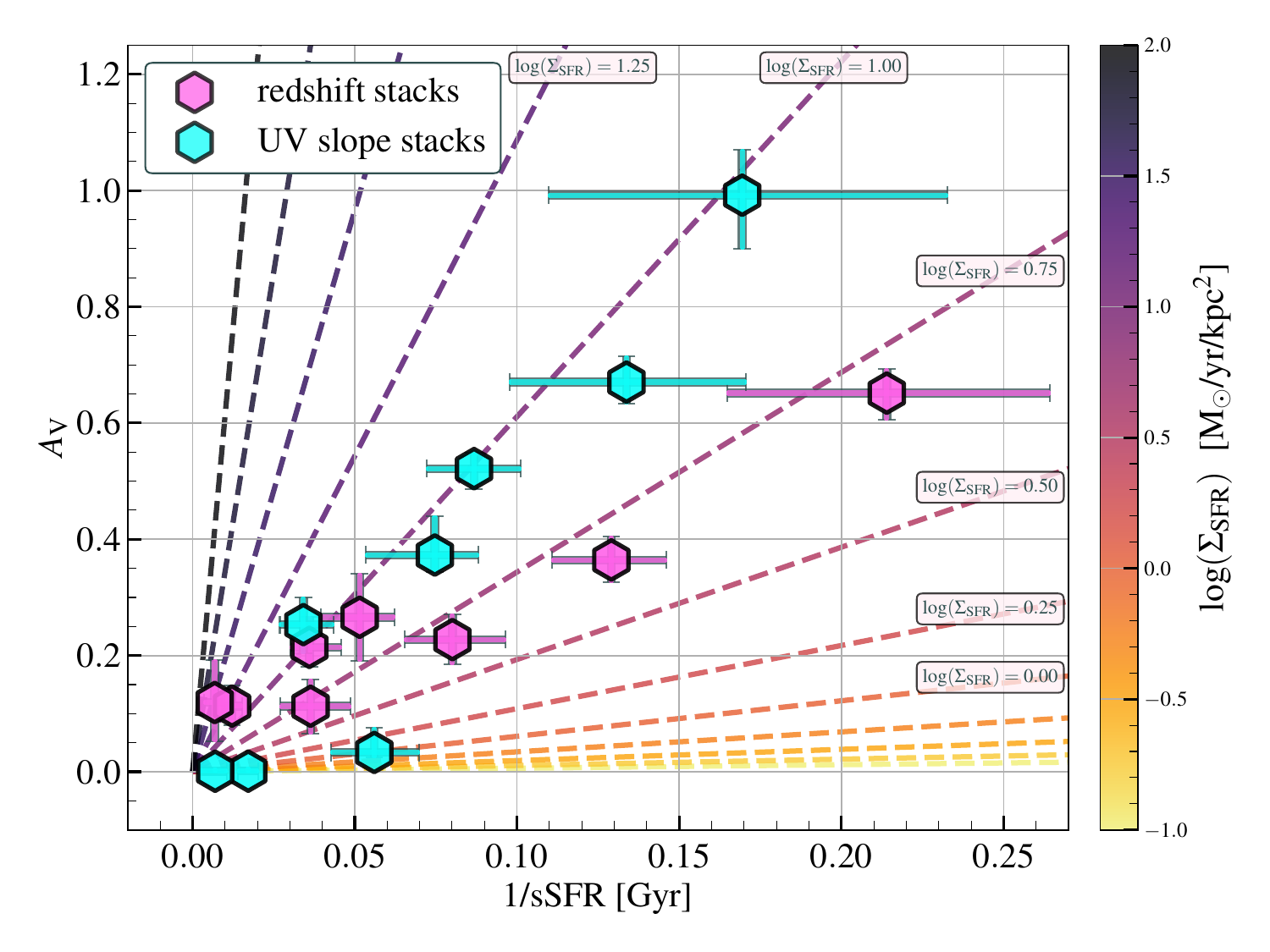}
\caption{
Strong anti-correlation between the dust attenuation ($A_{\rm V}$) and specific star formation rate (sSFR) for average galaxies at $3 < z_{\rm spec} < 14$. The pink and cyan data points represent the measurements based on 631 NIRSpec prism spectra stacked in 8 redshift and 8 UV slope bins. For the simple high-redshift galaxies the x-axis can be interpreted as a star formation timescale, with 1/sSFR closely tracing the time required to form the observed stellar mass. The appearance of significant attenuation as early as $\sim 30$ Myr indicates rapid dust formation, favoring dust production by core-collapse supernovae. We model the $A_{\rm V}$--sSFR anti-correlation analytically, assuming that all dust is produced by supernovae. This results in a linear scaling relation between the dust mass and stellar mass. Dust mass is then converted to attenuation by assuming a dust distribution geometry. We predict a linear $A_{\rm V}$--1/sSFR relation at a fixed SFR surface density ($\Sigma_{\rm SFR}$), with the more compact galaxies appearing more attenuated; this is shown by the colored dashed lines. The measurements for the stacked spectra are consistent with predictions for $\log(\Sigma_{\rm SFR}) \sim 0.5-1.0$ $M_{\odot}\,{\rm yr}^{-1}{\rm kpc}^{-2}$ galaxies. 
}
\label{fig: stacks}
\end{figure*}

Figure \ref{fig: stacks} shows the measured V-band dust attenuation ($A_{\rm V}$) and specific star formation rate (sSFR\footnote{sSFR = SFR/$M_{\star}$ where SFR is the star formation rate and $M_{\star}$ is the stellar mass.}) for average galaxies at $3 < z_{\rm spec} < 14$, exhibiting a strong empirical $A_{\rm V}$--sSFR anti-correlation. These measurements are based on uniformly-reduced JWST \citep{2006SSRv..123..485G} NIRSpec multi-shutter assembly \citep{2022A&A...661A..80J} prism spectra of 631 galaxies, stacked in 8 redshift and 8 UV slope bins \citep[][Methods]{Langeroodi24}. The compilation of this large dataset has been enabled by the unprecedented combination of depth and spectral coverage of NIRSpec \citep{2022A&A...661A..80J}, and the efficiency of its multi-shutter assembly in targeting numerous objects in a single exposure. The spectra were obtained through several programs \cite{2023ApJ...951L..22A, 2023Natur.622..707A, 2022arXiv221204026B, 2023arXiv230602467B, 2024arXiv240406531D, 2023arXiv230602465E, 2023arXiv231012340E, 2023ApJ...946L..13F, 2023ApJ...949L..25F, 2023arXiv230811609F, 2023arXiv230805735F, 2023arXiv230503042H, 2023ApJ...957L...7K, ZEIGHT, 2023MNRAS.tmp.3386L, 2023ApJ...947L..24M, 2023Natur.618..480R, 2023MNRAS.526.1657T, 2023Sci...380..416W} during the first and second JWST cycles. 

We detect significant attenuation for galaxies with ${\rm 1/sSFR} > 30$ Myr, indicating that dust is produced rapidly in the early universe. This is because 1/sSFR is a proxy for the age of galaxies (Methods). Specifically, $1/{\rm sSFR} = {\rm age}/f_{\rm SFR}$, where $f_{\rm SFR}$ depends on the star formation histories as well as the adopted SFR indicator. By definition, $f_{\rm SFR} = 1$ for a constant star formation history. In bursty high-redshift galaxies, where the majority of stellar mass was built up in the past $< 100$ Myr and is captured by the commonly adopted SFR indicators \citep{Langeroodi24}, $f_{\rm SFR} \sim 1$. In these galaxies $1/{\rm sSFR}$ closely traces the age of the stellar population. Therefore, the appearance of dust attenuation as early as ${\rm 1/sSFR} \sim 30$ Myr indicates that dust formation follows star formation on short timescales.


The inferred $A_{\rm V}$--1/sSFR correlation is not a product of systematic biases in the $A_{\rm V}$ and sSFR measurements (Methods). Rather, the dominant systematic biases affect the $A_{\rm V}$--1/sSFR distribution in a direction that is perpendicular to the observed trend. The SFRs of the stacked spectra are measured from their Balmer emission lines. The measured sSFRs are most accurately represented by the equivalent widths of these lines. These lines fall at rest-optical wavelengths, where the effect of dust reddening is relatively small. Furthermore, over/under-estimating the attenuation leads to over/under-estimating the line equivalent widths and hence the sSFRs. This would result in a systematic effect that is perpendicular to the observed $A_{\rm V}$--1/sSFR correlation.

Early-universe dust production is believed to be primarily driven by core-collapse supernovae \citep{1998ApJ...501..643D, Todini01, 2003MNRAS.343..427M, 2007ApJ...662..927D, 2011A&A...528A..14G, 2011A&A...528A..13G, 2011A&ARv..19...43G, 2014ApJ...788L..30D, Lesniewska19, 2020MNRAS.493.2490T, Ferrara22a}. We model the observed $A_{\rm V}$--1/sSFR trend assuming that dust is produced rapidly in the ejecta of core-collapse supernovae. This results in a dust mass that is effectively proportional to the number of supernovae, which in turn is proportional to the integrated star formation rate, or the stellar mass. In other words, dust production is primarily driven by the stellar mass buildup and star formation process. This $M_{\rm dust}$--$M_{\star}$ linear relation gives (Methods)
\begin{equation}
A_{\rm V} = \Upsilon \; \frac{\Sigma_{\rm SFR}}{\rm sSFR} \;,
\label{eq: main-text}
\end{equation}
where $\Sigma_{\rm SFR}$ is the star formation rate (SFR) surface density and $\Upsilon$ is a proportionality constant. The model predictions for various SFR surface densities are shown as the colored dashed lines in Figure \ref{fig: stacks}. The predicted linear $A_{\rm V}$--1/sSFR correlation is in close agreement with the measurements for the redshift and UV slope stacks.

\begin{figure*}
\centering\includegraphics[width = 1.0 \textwidth]{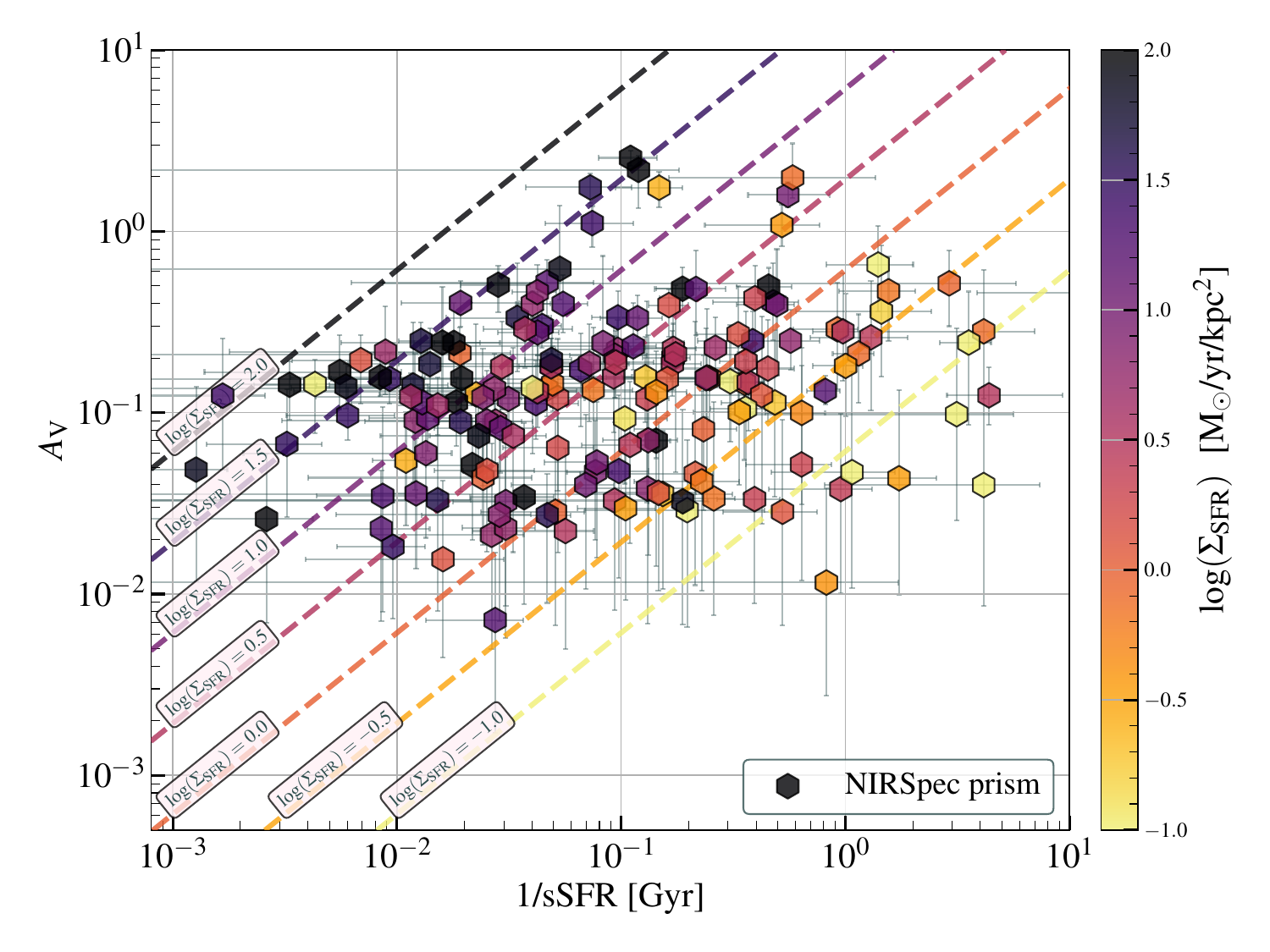}
\caption{
Dust attenuation ($A_{\rm V}$), specific SFR (sSFR), and SFR surface density ($\Sigma_{\rm SFR}$) for a subset of $3 < z < 13$ galaxies in our sample where measurements for all three parameters are available. We use this data to empirically infer the proportionality constant ($\Upsilon$) of our analytical model, $A_{\rm V} = \Upsilon \; \Sigma_{\rm SFR}/{\rm sSFR}$. Among other factors, $\Upsilon$ depends on the dust grains' size and chemical composition, the efficiency of dust destruction by supernovae forward and reverse shocks, and the efficiency of dust ejection by feedback-driven outflows. We find a best-fit value of $\Upsilon = 0.67^{+0.2}_{-0.2}\;{\rm cm}^2 {\rm g}^{-1}$, which is used for plotting the model prediction dashed lines in this diagram and Figure \ref{fig: stacks}. The inferred $\Upsilon$ is $\sim 30$ times lower than expectations for Milky Way-like grain properties, no supernovae dust destruction, and no feedback-driven dust ejection. The expected $\Upsilon$ can be lowered by $\sim 1$ dex if the early-universe dust is mostly comprised of large silicate grains (see Figure \ref{fig: grains}). This suggests that the role of supernovae dust destruction and dust ejection by feedback-driven outflows cannot be ignored. 
} 
\label{fig: individuals}
\end{figure*}

The dependence of $A_{\rm V}$ on the SFR surface density in Equation \ref{eq: main-text} indicates that more compact (i.e., higher $\Sigma_{\rm SFR}$) galaxies should appear more attenuated on average. In other words, at a fixed stellar mass and SFR we expect the smaller galaxies to appear more attenuated. As shown in Figure \ref{fig: individuals}, this trend is indeed seen for the subset of individual galaxies from our sample for which $A_{\rm V}$, sSFR, and SFR surface density measurements are available \citep{Langeroodi23}.

The proportionality constant ($\Upsilon$) in Equation \ref{eq: main-text} depends on a range of physical effects, including the efficiency of supernovae in producing dust, which itself depends on the dust formed per supernova and the stellar initial mass function (IMF); the fraction of dust that is destroyed by supernovae reverse and forward shocks, and through other mechanisms in the ISM; the fraction of dust that is removed from the galaxy by feedback-driven outflows; the dust distribution geometry; and the dust grains properties, including the distribution of their composition and size (Methods). We fit $\Upsilon$ directly against the $A_{\rm V}$, sSFR, and $\Sigma_{\rm SFR}$ distribution of individual $z > 3$ galaxies plotted in Figure \ref{fig: individuals} (see Methods for the details of the fit). We find $\Upsilon = 0.67^{+0.2}_{-0.2}\;{\rm cm}^2 {\rm g}^{-1}$. This is the proportionality constant adopted to generate the model $A_{\rm V}$--1/sSFR lines in Figures \ref{fig: stacks} and \ref{fig: individuals}. The measured $A_{\rm V}$ and sSFR of the redshift and UV slope stacks closely follow the trend predicted for $\log(\Sigma_{\rm SFR}) \sim 0.5$--1.0 $M_{\odot}\,{\rm yr}^{-1}{\rm kpc}^{-2}$ galaxies. 

Simple model assumptions result in an $\Upsilon$ value that is much higher than that measured here. We adopt empirically calibrated progenitor-mass-dependent supernovae dust yields, a Chabrier IMF, no supernovae or ISM dust destruction, no feedback-driven dust removal, a spherical shell dust distribution, and Milky Way-like dust grain properties. These assumptions yield $\Upsilon = 21.6\;{\rm cm}^2 {\rm g}^{-1}$ (Methods), a value which is $> 30$ times higher than inferred above   ($0.67^{+0.2}_{-0.2}\;{\rm cm}^2 {\rm g}^{-1}$). In other words, while Equation \ref{eq: main-text} successfully predicts the general linear $A_{\rm V}$--1/sSFR trend, high-redshift galaxies appear much less dusty than expected based on these simple assumptions. Below, we discuss these assumptions in turn.

The predicted $\Upsilon = 21.6\;{\rm cm}^2 {\rm g}^{-1}$ value is based on an assumed spherical shell dust distribution geometry. Other dust distribution geometries can result in different $\Upsilon$ values. For instance, a sphere of dust of the same radius results in an $\Upsilon$ value that is 3 times higher. Generally, a spherical distribution minimizes $\Upsilon$ and a modification of the dust distribution is unlikely to lower it substantially (Methods). The assumed Chabrier IMF also minimizes $\Upsilon$. This is because the high-redshift IMF is often assumed to be more top-heavy than the Chabrier IMF, which results in a higher supernova dust production per stellar mass formed \citep{Gall+2018}.

\begin{figure*}
\centering\includegraphics[width = 1.0 \textwidth]{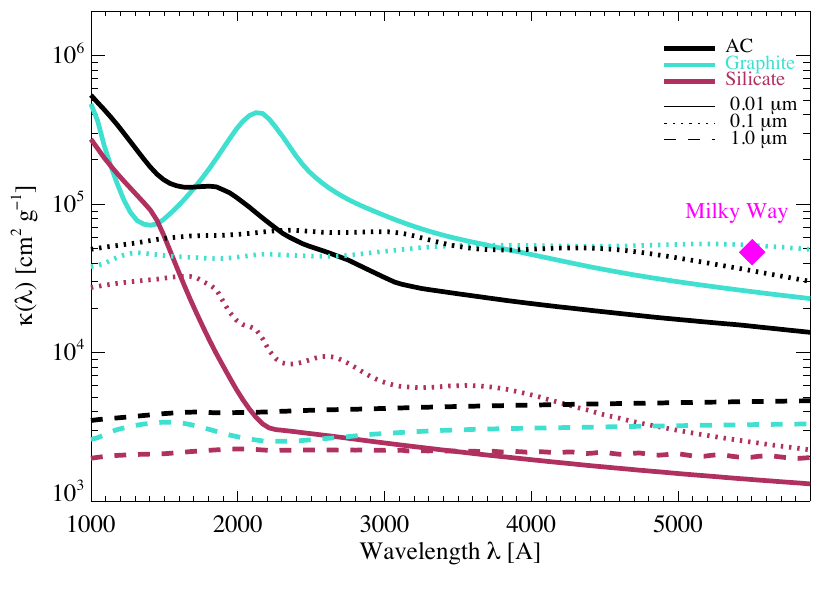}
\caption{
Wavelength dependence of the dust mass absorption coefficient ($\kappa_{\lambda}$) for various dust grain sizes and chemical composition based on theoretical models and observational constraints \citep{1991ApJ...377..526R, Zubko04, 2001ApJ...554..778L}. The $\kappa_{\lambda}$ values for the amorphous carbon (AC), graphite, and silicates dust grains are shown as the black, cyan, and red lines, respectively. For each chemical composition, $\kappa_{\lambda}$ for 0.01, 0.1, and $1.0 \mu$m dust grains are shown as the solid, dotted, and dashed lines, respectively. The V-band dust mass absorption coefficient ($\kappa_{\rm V}$) of Milky Way is indicated by the pink diamond. Larger and/or silicate dust grains result in $\kappa_{\rm V}$ values as much as $\sim 1$ dex lower than that of the Milky Way. We find that the early-universe galaxies appear $\sim 30$ times less attenuated than if their dust content was mostly comprised of Milky Way-like grains. This indicates that early-universe dust is dominated by large ($> 0.1 \mu$m) grains with silicate compositions. These features are indeed expected to be characteristic of newly formed dust in the ejecta of core-collapse supernovae. 
} 
\label{fig: grains}
\end{figure*}

Milky Way dust grain chemical compositions and sizes are unlikely to be representative of dust grains in high-redshift galaxies \citep{2014ApJ...788L..30D}. Indeed, the attenuation curves of star-forming regions significantly deviate from the average Milky Way extinction curve \citep{1991ApJ...374..580B,2014MNRAS.445...93D,2016MNRAS.455.4373D,2020MNRAS.492.3779H,2024arXiv240205996M,2024arXiv240710760L,2024arXiv240508445R}. As shown in Figure \ref{fig: grains}, different grain properties can result in reddening curves that are markedly different from the average Milky Way curve. The dust grains properties enter our estimate of $\Upsilon$ via the V-band dust mass absorption coefficient ($\kappa_{\rm V}$), where $\Upsilon$ scales linearly with $\kappa_{\rm V}$ (Methods). Figure \ref{fig: grains} shows that $\kappa_{\rm V}$ can be lower than the Milky Way value by $\sim 1$ dex, in particular if the high-redshift dust is mostly characterized by grains much larger than $0.1\;\mu$m and/or with silicate compositions. Such grains are expected for freshly formed dust in  supernova ejecta \citep{2014Natur.511..326G, 2020MNRAS.492.3779H}. This modification can bring the predicted $\Upsilon$ close to the observed value. It would imply that dust in high-redshift galaxies is dominated by large grains and/or silicate dust \citep{2020MNRAS.492.3779H}. However, since our assumptions for the dust distribution geometry and IMF minimize the $\Upsilon$, modifying the grain properties alone is not enough to explain the low empirically measured $\Upsilon$. 


The fraction of supernova-generated dust that survives the supernova shocks is uncertain \cite{2018SSRv..214...53M, Schneider+2024}. Most models agree that core-collapse supernovae must dominate the high-redshift dust production budget. This is due to their rapid dust production as well as a much higher dust yield per formed stellar mass compared to the other proposed dust production channels \citep{Marassi19}. However, theoretical models suggest that the reverse supernovae shocks can destroy part of the supernova-generated dust on $\gtrsim 10,000$ yr timescales. Moreover, the forward supernova shocks are also thought to destroy dust that is already present in the ISM \citep{2020ApJ...902..135S}. Thus, in principle, the net contribution of supernovae to the ISM dust may be lower than the total amount of dust they produce. However, in practice, as the efficiency of shocks in destroying dust is proportional to the dust-to-gas ratio in the ISM \citep{2020ApJ...902..135S}, at high redshifts where the ISM dust density is low, supernovae are net dust producers.

Alternatively, dust removal can play a role in regulating the dust content of high-redshift galaxies. The $1/{\rm sSFR} < 30$ Myr stacks in Figure \ref{fig: stacks} exhibit negligible attenuation. This is consistent with recent theoretical work suggesting that such extreme starbursts lead to super-Eddington radiation feedback which efficiently removes the gas and dust from the system. The resulting dust-free galaxies are expected to have enhanced UV flux. This mechanism is proposed to explain the lack of evolution in the bright end of the UV luminosity function at $z > 10$ \citep{Ferrara23a, Ferrara24}. Unfortunately, the low 1/sSFR end of our stacked data (Figure \ref{fig: stacks}) does not have enough signal to support fitting a feedback dust removal model, especially since its effect might be non-linear, i.e., more pronounced at the high sSFRs and not important at low sSFRs. 

In our data the effect of dust destruction is fully degenerate with the dust removal and our observations are only sensitive to the combined effect. Given the small empirically measured proportionality constant ($\Upsilon$), if we assume that the V-band dust mass absorption coefficient ($\kappa_{\rm V}$) for high-redshift galaxies is 10 times lower than the Milky Way value, $\sim 70\%$ of supernova-generated dust must have been either destroyed or ejected by feedback-driven outflows.

The linearity of the observed $A_{\rm V}$--1/sSFR correlation further confirms that dust builds up rapidly at the probed redshifts ($z > 3$). This is because the predicted linear $A_{\rm V}$--1/sSFR trend is the product of the linear $M_{\rm dust}$--$M_{\star}$ scaling relation, which requires the dust buildup to immediately follow the stellar mass buildup. Slow dust production channels lead to non-linear $M_{\rm dust}$--$M_{\star}$ and consequently non-linear $A_{\rm V}$--1/sSFR relations in the early-universe since there will be a significant delay between the epoch of star formation and the appearance of significant dust. Equation \ref{eq: main-text} was derived with the instantaneous recycling approximation, which assumes that the entire supernovae dust yield is immediately added to the system. This assumption breaks down for stellar populations younger than $\sim 30$ Myr, as there is not enough time for all the progenitor masses to burst in supernovae. This can lead to deviations from a linear $A_{\rm V}$--1/sSFR correlation, but only at ${\rm 1/sSFR} < 30$ Myr where we observe negligible attenuation. For the other stacks, the instantaneous recycling is an accurate approximation. 

Dust formation, destruction, and removal mechanisms that scale non-linearly with the stellar mass should appear as deviations from the straight lines in Figure \ref{fig: stacks}. One such scenario is ISM grain growth, where the dust mass not only scales with the stellar mass (since the gas mass scales with the stellar mass), but also with the age of the galaxy. Figure \ref{fig: stacks} shows no strong evidence for non-linear contributions. This suggests that if ISM grain growth significantly contributes to the total dust budget, it only plays a role at lower redshifts and lower sSFRs than those covered in this study \citep{2020MNRAS.493.2490T}. At lower redshifts, where the potential contribution from evolved AGB stars and the ISM grain growth channels become significant, the excess dust buildup should cause upward deviations from the straight $A_{\rm V}$--1/sSFR trends. 


\clearpage

\clearpage
\section{Methods}\label{sec: methods}

\subsection{Data} \label{sec: data}

The NIRSpec multi-shutter assembly \citep[MSA;][]{2022A&A...661A..80J, ferruit+2022} prism spectra of the galaxies in our sample are reduced as detailed in \citet{Langeroodi23, Langeroodi24}, and stacked in 8 redshift and 8 UV slope bins as detailed in \citet{Langeroodi24}. For the stacked spectra, we adopt the spectral energy distribution (SED) fits and the corresponding stellar population inferences from \citet{Langeroodi24}. For the individual galaxies, we adopt the SED and morphological fits from \citet{Langeroodi23}. Below, we provide an overview of these fits.

The stacked spectra were fitted with a SED-fitting code specifically designed to handle the unknown spectral resolution of the stacks, where the spectral features are significantly broadened due to spectroscopic redshift inaccuracies. This is achived by treating the spectral resolution as a free parameter. The star formation histories were fitted non-parametrically. The first temporal bin spans the first $10^{6.5}$ yr in lookback time, while the rest are evenly spaced in 0.5 dex intervals out to the earliest onset of star formation, assumed to be $z = 20$. A continuity prior was enforced to penalize against sharp SFR transitions in adjacent temporal bins \citep{2019ApJ...876....3L}. The dust attenuation was modeled with a \citet{2000ApJ...533..682C} curve. SFRs were measured from the Balmer lines, after correction for dust attenuation. These lines were fitted simultaneously while fitting the continuum. The inferred sSFR and $A_{\rm V}$ are shown in Figure \ref{fig: stacks}.

From the sample presented in \citet{Langeroodi23}, a subset of 173 galaxies at $4 < z < 10$ have both SED and morphology fits. The stellar population of each galaxy is modeled by simultaneously fitting its available HST photometry, NIRCam photometry, and NIRSpec MSA prism spectroscopy with \texttt{prospector} \citep{Johnson21}. The adopted SED-fitting setup was similar to that used to fit the stacked spectra \cite{Langeroodi24}. The star formation history was modeled non-parametrically with a continuity prior. The dust attenuation was modeled with the \citet{2013ApJ...775L..16K} generalization of the \citet{2000ApJ...533..682C} curve. SFRs were measured from the Balmer lines (after correction for attenuation), which were fitted using \texttt{pPXF} \citep{pPXF1, pPXF2, pPXF3}. The inferred sSFR and $A_V$ for these galaxies are presented in Figure \ref{fig: individuals}. 

The data points in Figure \ref{fig: individuals} are color-coded with their corresponding SFR surface densities. We measure the SFR surface densities as $\Sigma_{\rm SFR} = {\rm SFR}/(2 \pi R_{e}^2)$, where $R_{e}$ [kpc] is the half-light radius. The half-light radii were measured in \citet{Langeroodi23} by fitting the NIRCam photometry using the \texttt{galight} software \citep{galight}, a wrapper around the \texttt{lenstronomy} image modeling tool \citep{lenstronomy1, lenstronomy2, lenstronomy3}. For each galaxy, the fit was performed in the NIRCam short-wavelength image corresponding to its rest-frame 2000\AA\ emission. All the bright sources in a $2'' \times 2''$ image cutout were identified and simultaneously fitted with Sersic profiles \citep{1968adga.book.....S} after convolution with a point spread function (PSF). The PSFs were constructed using \texttt{WebbPSF} \citep{wPSF2, wPSF3}. 

We note that the dust attenuation of the stacks and individual galaxies could also be inferred from their Balmer decrements. However, as discussed in \citet{Langeroodi24}, the attenuation values inferred by SED-fitting are expected to be more accurate and provide a much wider redshift coverage. At $z < 5$ for the stacks and $z < 4$ for individual galaxies H$\beta$ cannot be resolved from the [OIII]$\lambda\lambda$4959,5007\AA\ doublet in prism spectroscopy. Moreover, at $z > 7$ H$\alpha$ gets redshifted out of the NIRSpec prism coverage. Other Balmer lines in NIRSpec prism are typically weak, blended with other lines, and do not provide a long enough wavelength baseline for accurate attenuation measurements. As a result, SED-fitting dust attenuation measurement enables a much wider redshift coverage compared to what is possible through Balmer decrements. Furthermore, recent studies have shown that significant deviations from the Case B recombination's intrinsic Balmer line ratios might be common at high redshifts \citep{2024arXiv240320118Y, 2024arXiv240515859M}. This can result in significant systematics in the attenuation values measured through the observed Balmer decrements. 

\subsection{Degeneracies} \label{sec: degeneracies}

\begin{figure*}
\centering\includegraphics[width = 0.9 \textwidth]{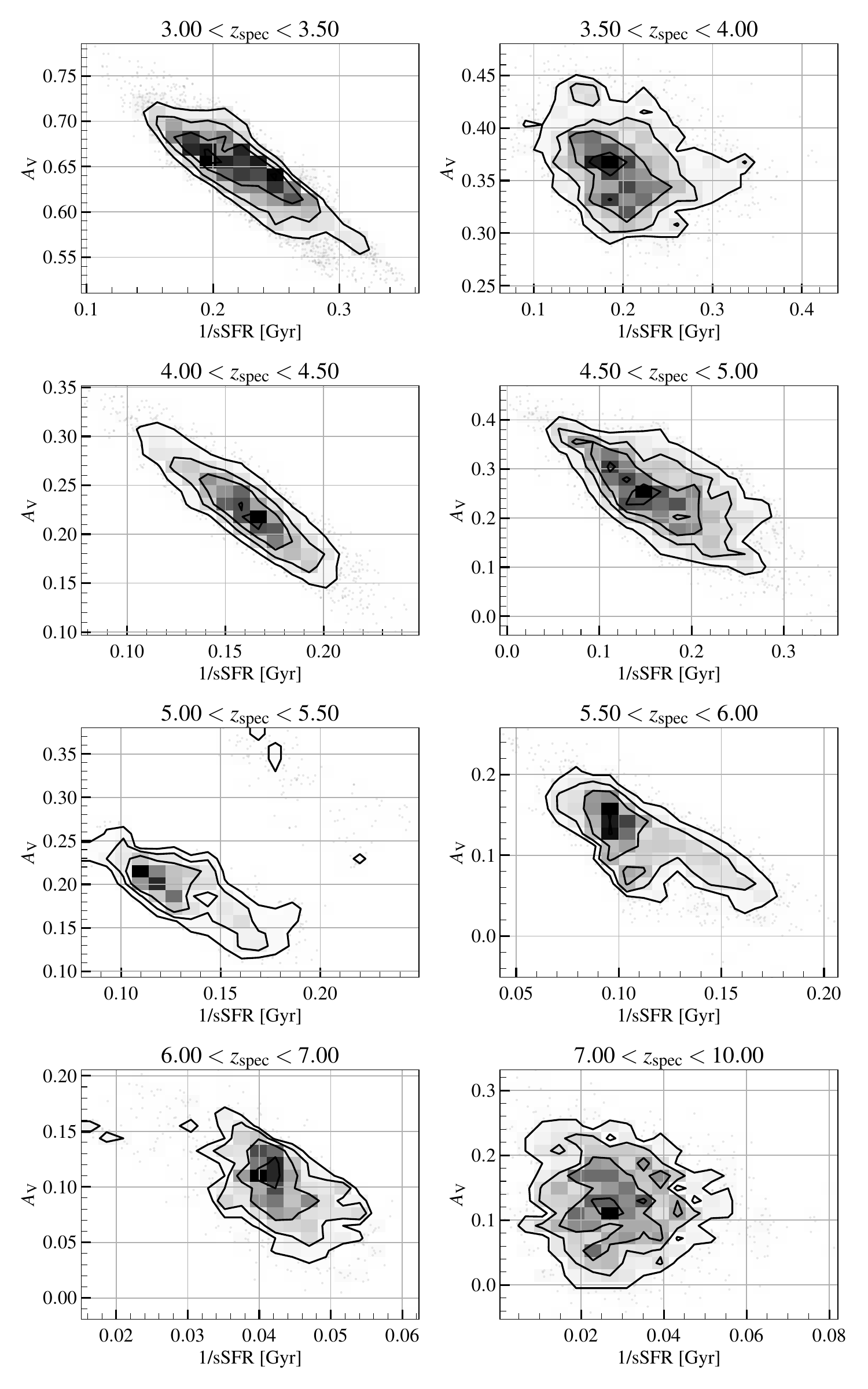}
\caption{Marginalized posterior distribution of the measured $A_{\rm V}$ and 1/sSFR for the 8 redshift stacks, if SFRs are measured through SED-fitting. The redshift range of each stack is indicated on the top of its corresponding panel.} 
\label{fig: posterior redshift SED}
\end{figure*}

\begin{figure*}
\centering\includegraphics[width = 0.9 \textwidth]{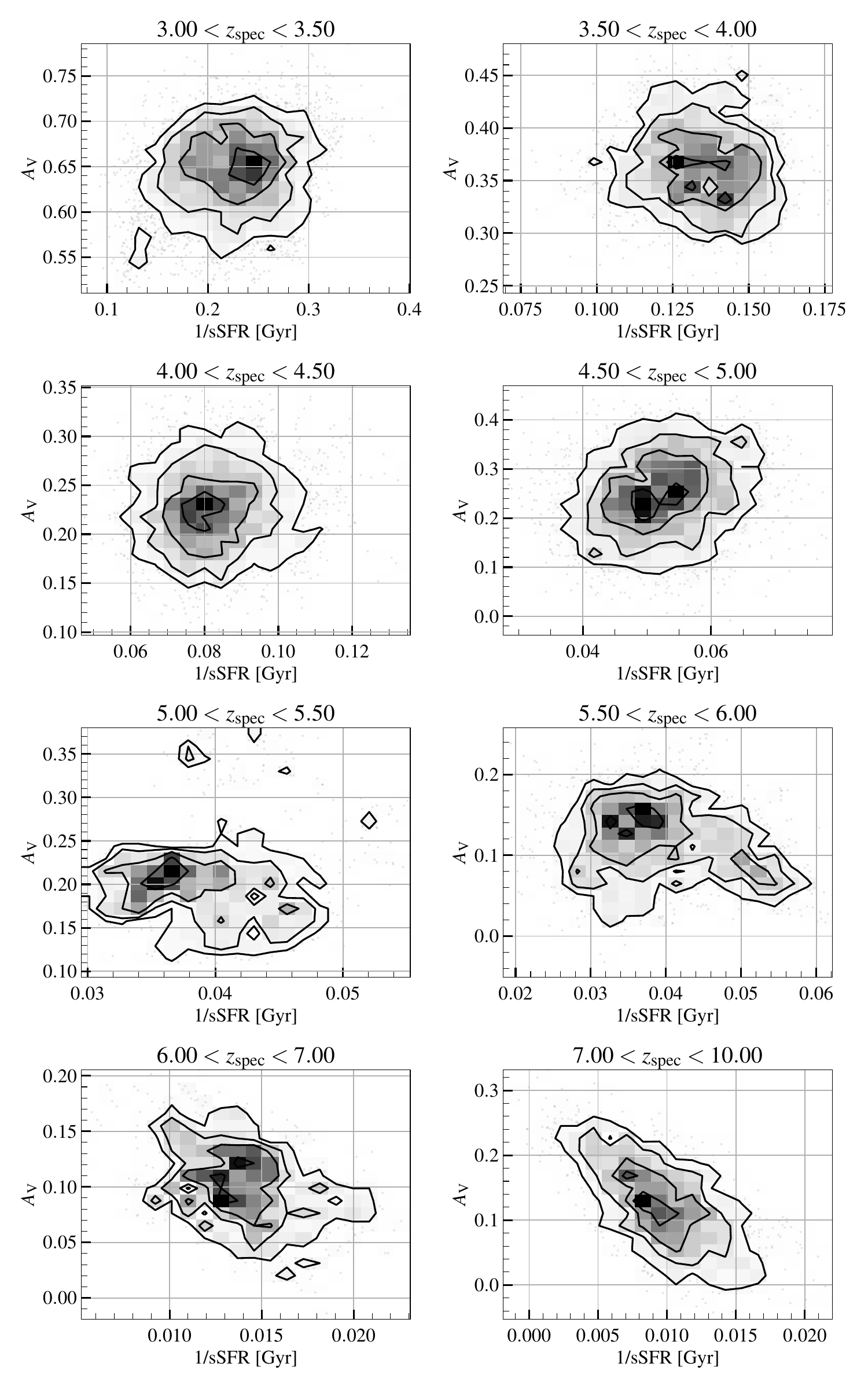}
\caption{Marginalized posterior distribution of the measured $A_{\rm V}$ and 1/sSFR for the 8 redshift stacks, if SFRs are measured from the Balmer emission lines. The redshift range of each stack is indicated on the top of its corresponding panel.} 
\label{fig: posterior redshift lines}
\end{figure*}

\begin{figure*}
\centering\includegraphics[width = 0.9 \textwidth]{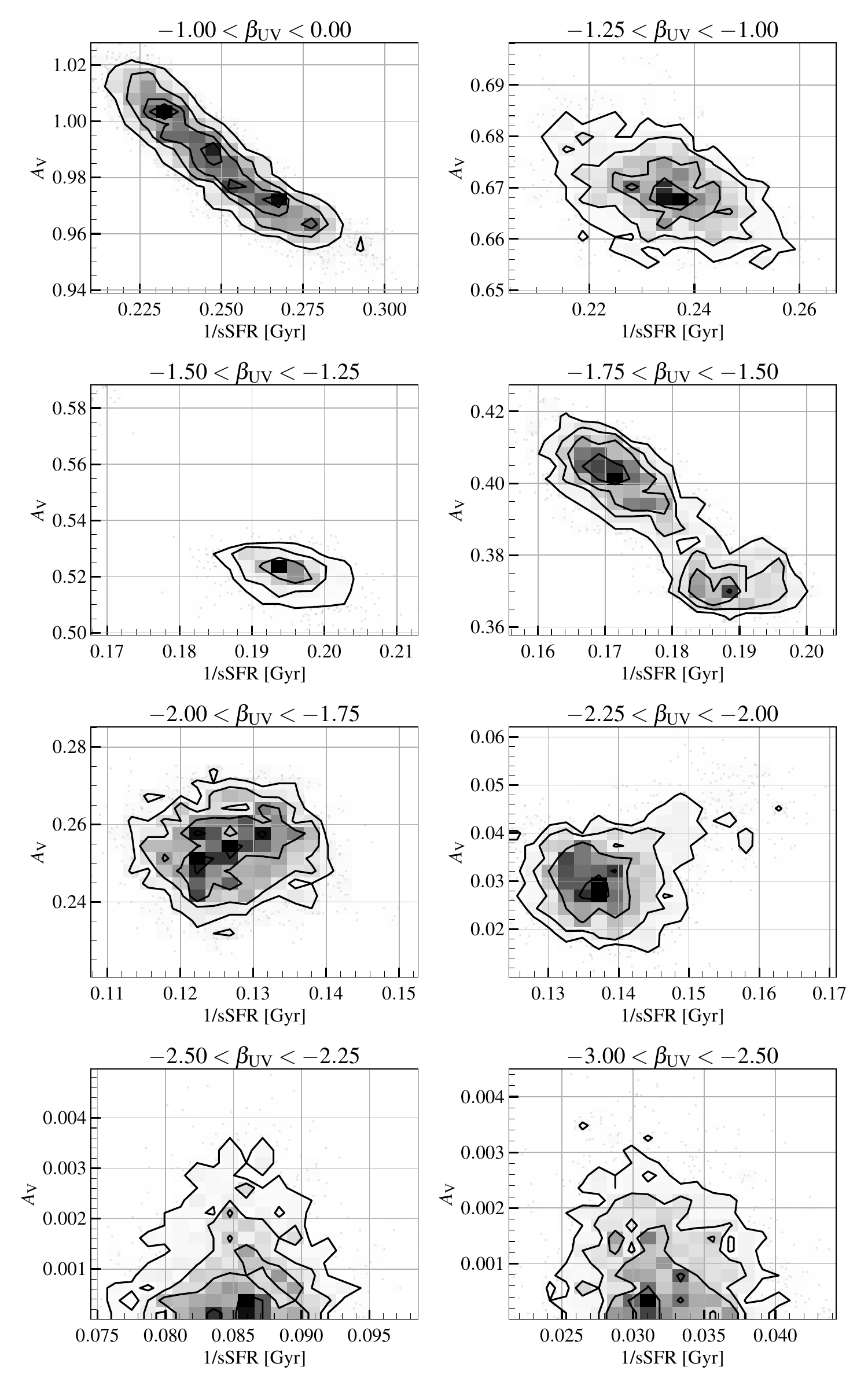}
\caption{Marginalized posterior distribution of the measured $A_{\rm V}$ and 1/sSFR for the 8 UV slope ($\beta_{\rm UV}$) stacks, if SFRs are measured through SED-fitting. The UV slope range of each stack is indicated on the top of its corresponding panel.} 
\label{fig: posterior uvslope SED}
\end{figure*}

\begin{figure*}
\centering\includegraphics[width = 0.9 \textwidth]{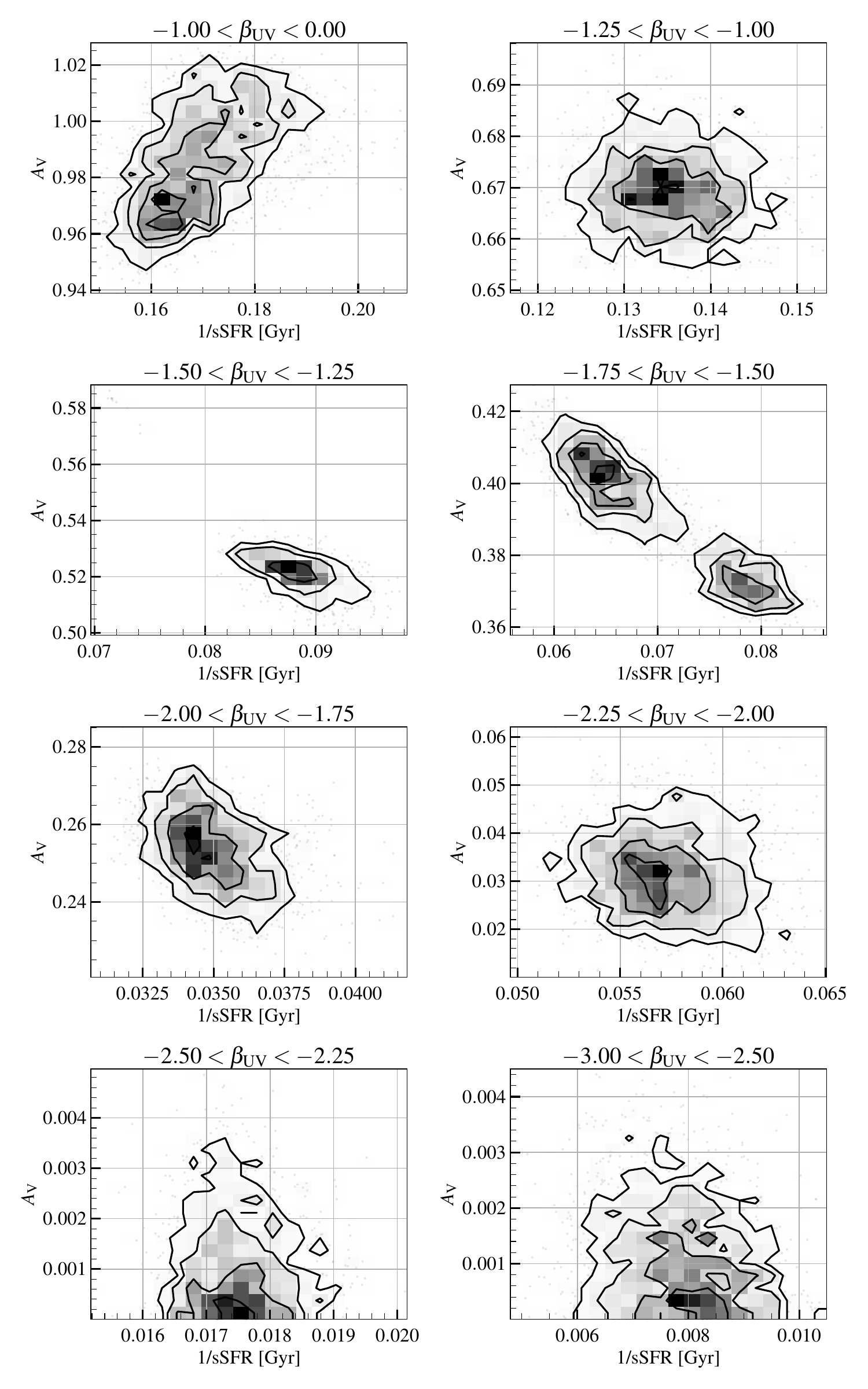}
\caption{Marginalized posterior distribution of the measured $A_{\rm V}$ and 1/sSFR for the 8 UV slope ($\beta_{\rm UV}$) stacks, if SFRs are measured from the Balmer emission lines. The UV slope range of each stack is indicated on the top of its corresponding panel.} 
\label{fig: posterior uvslope lines}
\end{figure*}

Here, we establish that the $A_{\rm V}$--1/sSFR trend observed for the stacked spectra (Figure \ref{fig: stacks}) is not the product of a degeneracy between the inferred $A_{\rm V}$ and sSFR. For this purpose, we investigate the marginalized posterior probability distribution of the inferred $A_{\rm V}$ and sSFR for each stack, based on the dynamic nested sampling chain of its SED fit \citep{Langeroodi24}. The posterior distributions are shown in Figures \ref{fig: posterior redshift SED}, \ref{fig: posterior redshift lines}, \ref{fig: posterior uvslope SED}, and \ref{fig: posterior uvslope lines}. For completeness, we present the posterior distributions for the case where the SFRs are measured by SED-fitting (Figures \ref{fig: posterior redshift SED} and \ref{fig: posterior uvslope SED}) as well as the case where the SFRs are measured from Balmer lines (Figures \ref{fig: posterior redshift lines} and \ref{fig: posterior uvslope lines}). These figures confirm that the degeneracy between the measured $A_{\rm V}$ and sSFR is either negligible, or in a direction that is perpendicular to the observed trend. 

\subsection{Relation between time and specific star formation} \label{sec: ssfr}

Here, we establish that 1/sSFR can be adopted as a proxy for the age of young galaxies. This is particularly important, as it motivates interpreting the observed linear $A_{\rm V}$--1/sSFR correlation for average galaxies (Figure \ref{fig: stacks}) as a temporal buildup of dust in the early-universe. 

We start by prescribing a general form of the star formation history (SFH) evolution with lookback time, $t_{\ell}$, i.e. the difference between the age of the Universe at the redshift of the source, $z_s$, and that at redshift, $z$, at which the SFR is computed,
\begin{equation}\label{eq:tl}
    t_{\ell}(z) = t_H \int_{z_s}^z \frac{dz'}{(1+z')E(z')},
\end{equation}
where $E(z) = [\Omega_m(1+z)^3+\Omega_\Lambda)]^{1/2}$. We adopt the widely used ``$\tau$-model'' in which the SFR exponentially increases with cosmic time, 
\begin{equation}\label{eq:SFH}
    {\rm SFR}(t_{\ell}) = {\rm SFR_{\rm obs}}\, e^{-t_{\ell}/\tau},
\end{equation}
where $\tau$ is a characteristic $e$-folding time. The $\tau$-model fairly well describes the SFH of early galaxies \citep[see e.g.][]{Pallottini22}, and is implemented in most SED fitting codes (e.g. \texttt{beagle} \citep{Chevallard16};  \texttt{bagpipes} \citep{Carnall18}; \texttt{prospector} \citep{Johnson21}). 

Using Equation \ref{eq:SFH} we can compute the total stellar mass at $t_{\ell}=0$ from the beginning of star formation at $z=z_\star$ 
\begin{equation}\label{eq:Ms}
    M_\star(t_{\ell}=0) = {\rm SFR_{\rm obs}}\, \int_0^{z_\star} e^{-t_{\ell}/\tau}\ dt_{\ell} = {\rm SFR_{\rm obs}}\,\tau;
\end{equation}
in practice, $M_\star$ is insensitive to the value of $z_\star$. Thus, we have taken $z_\star = \infty$. Using Equations \ref{eq:SFH} and \ref{eq:Ms}, it follows that, at $t_{\ell}=0$  
\begin{equation}\label{eq:sSFR}
    {\rm sSFR} = \frac{{\rm SFR_{\rm obs}}}{M_\star}= \frac{1}{\tau}.
\end{equation}

More generally, we can start from the the stellar mass ($M_{\star}$) for a general star formation history
\begin{equation}
    M_{\star} = \int_0^{t_{\rm form}} {\rm SFR}(t_\ell) dt_l\;,
\label{eq: M*}
\end{equation}
where $t_{\rm form}$ is the lookback time at the earliest epoch of star formation. To allow for a generic SFH, we define the average star formation rate as 
\begin{equation}
    \langle {\rm SFR} \rangle = \frac{\int_0^{t_{\rm form}} {\rm SFR}(t_\ell) dt_\ell}{\int_0^{t_{\rm form}} dt_\ell} \;,
\end{equation}
giving
\begin{equation}
    M_{\star} = \langle {\rm SFR} \rangle \, t_{\rm form}\;.
\end{equation}
We can use this to express sSFR in terms of $t_{\rm form}$
\begin{equation}
    {\rm sSFR} = \frac{\rm SFR_{obs}}{M_{\star}} = f_{\rm SFR} / t_{\rm form}\;,
\end{equation}
where $f_{\rm SFR} \equiv$ SFR$_{\rm obs}$/$\langle {\rm SFR} \rangle$ depends on the adopted SFR indicator used to measure SFR$_{\rm obs}$ and the deviation of star formation history from a constant SFR model. By definition, $f_{\rm SFR} = 1$ for a constant SFR model and $f_{\rm SFR} > 1$ for an exponentially rising star formation history. SFR indicators such as the UV continuum and Balmer lines trace the SFR over the past $\sim 100$ Myr and the past few $\sim 10$ Myr, respectively \citep[see e.g.,][and references therein]{2016ApJ...833..254S, Langeroodi24}. Hence, for young high-redshift galaxies where most of the stellar mass was built up in the past $< 100$ Myr \citep{Langeroodi24}, we have $f_{\rm SFR} \sim 1$. 

\subsection{Model} \label{sec: model}

In this Section we derive the relation between the V-band dust attenuation, $A_V$, and the specific star formation rate, $\rm sSFR = SFR_{\rm obs}/M_\star$ of a galaxy, where $\rm SFR_{\rm obs}$ and $M_\star$ are the observed star formation rate and stellar mass of the system, respectively. 

We note that at high redshift ($z \gtrsim 5$) the dust production is almost completely dominated by supernovae \citep{Todini01,Lesniewska19, Ferrara22a} as the timescales for other dust production channels like AGB and evolved stars are too long\footnote{We implicitly also make the Instantaneous Recycling Approximation (IRA).} to be relevant \citep{Marassi19}. In addition, dust growth proceeds too slow to be competitive with supernova dust production \cite{Ferrara16, Priestley21, Dayal22}. We conservatively assume that no dust is lost via outflows.  With this preamble, the dust mass can be written as 
\begin{equation}\label{eq:Md}
    M_d = y_d \nu M_\star = y_d \nu\ {\rm SFR_{\rm obs}}\,\tau,
\end{equation}
where $y_d$ is the net dust mass formed per supernova (i.e. after the processing of the newly formed dust in supernova ejecta by the reverse shock \citep{Bianchi07}), and $\nu$ is the number of supernovae per unit stellar mass formed, which depends on the stellar initial mass function (IMF). It is worth noting that Equation \ref{eq:Md} remains valid also in the case in which dust destruction and growth approximately balance\footnote{Using the dust production/destruction rates given in \citet{Dayal22}, this occurs when the growth timescale is approximately equal to the free-fall time scale of the gas.}. Deriving $\tau$ from Equation \ref{eq:Md} and substituting it into Equation \ref{eq:sSFR} we get
\begin{equation}\label{eq:sSFR1}
    {\rm sSFR} = \left(\frac{y_d\nu}{M_d}\right) {\rm SFR_{\rm obs}}.
\end{equation}

We can write the dust optical depth, $\tau_V$, in the (restframe) V-band ($5100$\AA), assuming that the dust is distributed on a scale comparable to the effective radius, $r_e$, of the galaxy. Then, following \citet{Ferrara24}, we write\footnote{Eq. \ref{eq:AV} assumes a dust shell. For an uniform, spherical distribution $A_V$ is 3 times larger.} 
\begin{equation}\label{eq:AV}
    A_{V} = 1.086 \tau_V =  1.086 \frac{\kappa_{V}}{4\pi r_e^2}  M_d,    
\end{equation}
where $\kappa_{V}$ is the dust mass absorption coefficient ($\rm cm^2\ g^{-1})$, which depends on the dust composition and its grain size distribution \citep[see Figure \ref{fig: grains};][]{1991ApJ...377..526R, 1996MNRAS.282.1321Z, 2001ApJ...554..778L}. 

By substituting the expression for $M_d$ in Equation \ref{eq:AV} into Equation \ref{eq:sSFR1}, and by further introducing the star formation surface density, $\Sigma_{\rm SFR} = {\rm SFR_{\rm obs}}/2\pi r_e^2$, we obtain the ${\rm sSFR}-A_V$ relation
\begin{equation}\label{eq:relation0}
{\rm sSFR} = 0.543\, {y_d\nu \kappa_V}\, \frac{\Sigma_{\rm SFR}}{A_V} \equiv \Upsilon \frac{\Sigma_{\rm SFR}}{A_V}.    
\end{equation}
This relation shows that $A_V$ is linearly correlated with $1/{\rm sSFR}$ with a slope coefficient $\Upsilon \Sigma_{\rm SFR}$, where $\Upsilon$ depends on the dust yield, IMF (via $\nu$), and the V-band optical properties of the grains. 

For reference, we adopt the following values motivated by the local Universe observations: $y_d \nu = 0.004\ M_{\odot}$, appropriate for a Chabrier IMF \citep{Gall+2018}; and $\kappa_{V} = 4.74 \times 10^4\ {\rm cm}^2 {\rm g}^{-1}$, corresponding to a Milky Way extinction curve. These values yield $\Upsilon = 21.6\ \rm cm^2\ g^{-1}$. Then, in standard units, Equation \ref{eq:relation0} reduces to
\begin{equation}\label{eq:relation}
    {\rm sSFR} = {21.6} \left(\frac{\Sigma_{\rm SFR}}{\rm M_{\odot} yr^{-1} kpc^{-2}}\right){A_V^{-1}}\quad \rm Gyr^{-1}.    
\end{equation}

\subsection{Generalization of the model}

In this Section, we generalize the model derivation of the previous Section to capture a more diverse array of dust destruction, dust removal, and dust distribution geometries. First we generalize our assumptions for the dust destruction and removal by rewriting the relation between dust mass and stellar mass as 
\begin{equation}
    M_d = \tilde{y}_d\, \nu\, f_{\rm des} \, f_{\rm rem}\, M_\star \;,
    \label{eq: gen-dust}
\end{equation}
where $\tilde{y}_d$ is the total dust mass formed per supernova (i.e. before the processing by supernova reverse and forward shocks); $\nu$ is the number of supernovae per unit stellar mass formed, which depends on the IMF; $f_{\rm des}$ is the mass fraction of formed dust that is preserved against destruction, most notably by the supernovae reverse shocks \citep{Bianchi07, Gall+2018}; and $f_{\rm rem}$ is the mass fraction of the generated dust that is retained against feedback-driven outflows. Dividing both sides of Equation \ref{eq: gen-dust} by the observed star formation rate (${\rm SFR}_{\rm obs}$) gives 
\begin{equation}
    {\rm sSFR} = \frac{\tilde{y}_d \, \nu \, f_{\rm des} \, f_{\rm rem}}{M_d} \, {\rm SFR_{obs}} \;.
    \label{eq: gen-sSFR}
\end{equation}

Now we generalize our assumption for the dust distribution geometry. In Section \ref{sec: model}, we converted the dust mass to attenuation by assuming a spherical shell distribution of dust at the half light radius of the galaxy. We can generalize this assumption as
\begin{equation}
    A_{\rm V} = 1.086 \frac{g \, \kappa_{\rm V}}{4 \pi r_e^2} M_d \;,
    \label{eq: gen-geometry}
\end{equation}
where $g$ captures the effect of different dust distribution geometries. For instance, a spherical shell at $r_e$ gives $g = 1$, while a sphere of radius $r_e$ gives $g = 3$. This generalization can capture more complex dust distribution geometries. For example, a power-law dust distribution ($M_d(r) \propto r^{-\alpha}$) results in $g = (3-\alpha)/(1-\alpha)$. In general, a shell-like geometry minimizes $g$. 

Using Equation \ref{eq: gen-geometry} to insert for $M_d$ in Equation \ref{eq: gen-sSFR}, we get 
\begin{equation}
    A_{\rm V} = 0.543\, \tilde{y}_d \, \nu \, g\, k_{\rm V}\, f_{\rm des} \, f_{\rm rem} \, \frac{\Sigma_{\rm SFR}}{\rm sSFR} \equiv \Upsilon \frac{\Sigma_{\rm SFR}}{\rm sSFR} \;.
\end{equation}
We emphasize that this general form of the $A_{\rm V}$--1/sSFR relation was derived without any explicit assumptions for the SFH, and as such is applicable to any galaxy regardless of the exact shape of its SFH. In the main text, we discuss the implications of the empirically constrained $\Upsilon$ for the $g$, $\kappa_{\rm V}$, $f_{\rm des}$, and $f_{\rm rem}$ of the early-universe galaxies. 

\subsection*{Data availability}

The raw JWST NIRSpec data used in this work can be accessed at the JWST archive (archive.stsci.edu). The stacked spectra are available upon request. 

\clearpage

\clearpage

\bibliography{main}

\end{document}